# Microstructure and correlated mechanical properties study of Ni-(Fe, Co)-Mn-(Al, In) as-spun ribbons

Chunyang ZHANG[1], Laureline PORCAR[2], Salvatore MIRAGLIA[2], Patricia DONNADIEU[1], Muriel BRACCINI[1], Richard HAETTEL[2] and Marc VERDIER[1]

[1]Univ. Grenoble Alpes, CNRS, Grenoble INP, SIMaP, F-38000 Grenoble, France

[2]Univ. Grenoble Alpes, CNRS, Grenoble INP, Institut Néel, F-38000 Grenoble, France

**Abstract:**

The Ni-Mn-based shape memory alloys as a promising candidate of elastocaloric material has been reported in many literatures, especially on bulk samples. The as-spun ribbon, which has a larger surface area and is more efficient for heat transfer, is rarely studied and hence of importance. In the present work, we succeeded in producing very long as-spun Ni-Fe-Mn-(Al, In) ribbons, with around 300 mm in length. The microstructure and mechanical properties of these as-spun ribbons were thoroughly investigated by scanning electron microscopy / electron backscattered diffraction (SEM/EBSD), nanoindentation and 3-points bending experiments. Through SEM/EBSD analyses, the microstructure and texture of the as-spun ribbons were studied. A gradient in microstructure exists along the thickness direction (TD) of the ribbon, which is induced by the temperature gradient during fast rate solidification, resulting in fine equiaxed grains along the surface contacted with the rotating wheel in melt spinning process and elongated grains, respectively. Both equiaxed and elongated grains possess a strong {001} fiber texture (<001>//TD). Nanoindentation analyses show little variation of hardness between the two different microstructures. The ductility index of both Ni-Fe-Mn-Al and Ni-Fe-Mn-In ribbons are within the range of intermetallic materials. The substitution of In by Al allows to increase very slightly the ductility index, which can reach 0.75. The fine equiaxed grains show better tensile resistance than the elongated grains in 3-points bending test. The substitution of In by Al improves the maximum bending strain by a factor of 3. The maximum strain for Ni-Fe-Mn-Al as-spun ribbons can reach 3 % before fracture. Fractography shows that the intergranular fracture is the main damage mechanism in these as-spun ribbons.

# 1. Introduction

Caloric effects described the temperature and entropy changes under the application of an external field to a given material [1-3]. The elastocaloric effect (eCE) is the mechanical analogue of the magnetocaloric effect and is related to the isothermal change of entropy or the adiabatic change of temperature that takes place when uniaxial stress is applied or released [4, 5]. Due to the priorities of energy savings and environmental protection, new refrigeration techniques based on the elastocaloric effect of solid state transformation induced by uniaxial stress have been proposed as potential alternatives to conventional cooling by vapor compression and have attracted significant attention in the literature. Most investigations in elastocaloric materials are focused on the shape memory alloys (SMAs), which exhibit ferroelastic phase transition, Cu-based SMAs [5, 6], Ni-Ti based SMAs [5, 7, 8] and Ni-Mn-based Heusler SMAs [9-12]. The eCE potentiality in materials have conducted to the development of elastocaloric cooling and heat pumping and devices based on compressive thermoelastic tubes, tensile sheet/ribbons and bending films [13-16]. Large eCE in these SMAs mainly originates from the large transformation entropy change $\Delta S_{tr}$ that represents the maximum value to be reached during isothermal stress-induced transition. There is a correlation between the structural entropy change and the volume differences of the two phases. Large elastocaloric effect can thus be expected in ferroelastic alloys with large volume change. Among SMAs, ferromagnetic shape memory alloys (FSMAs), where large magnetic-field induced strains was observed, attracted increased attention, owing to versatility and the possibility to used them in broad applications. However, while magnetic order is required for magnetocaloric properties, large volume changes and weak magnetic order is desirable for eCE [17]. All-d-metal Heusler shape memory alloys seem to answer this criterion [12]. In Ni-Mn-based FSMAs, large adiabatic temperature change can be achieved under relatively small stress [18] implying smaller fatigue upon stress cycling.

From the point of view of practical cooling applications, the combination of a high reversible caloric effect obtained in first-order solid-solid phase transition materials and excellent mechanical properties and stability is essential. However, the giant caloric effect is inseparable from a thermal hysteresis that results from the elastic incompatibility between the two solid phases. Generally, a cycling strain around 3% – 7% is obligatory for a fully strain induced martensitic transformation and the inverse transition, depending on composition, phase transition temperatures, microstructure, and experimental temperature *etc.* [19-21]. While, the poor mechanical properties (low ductility and brittle fracture) of Ni-Mn-based alloys restrict their use for applications when strain is applied. It is known that the ductility of ordered intermetallic alloys can be improved by alloying processes, rapid solidification and thermomechanical treatment [22]. By alloying of additional elements, *e.g.* B [10, 24], Co [23] and Fe [9] *etc.* in Ni-Mn-based alloys, the mechanical properties and eCE performance were largely increased.

Many investigations on stress-induced transformation and eCE effect were conducted on single crystals. However, polycrystalline materials are more important for technical applications due to their lower cost and ease of fabrication. Ribbon geometry is interesting because of a high surface-to-volume ratio, *i.e.* geometry favorable to a high heat exchange rate and cooling efficiency and to bypass the intrinsic fragility of intermetallic [25]. Melt spinning technique is a process with an extremely high cooling rate for the synthesis of Ni-Mn-based polycrystalline ribbons. However, the samples remain highly brittle due to the elongated textured polycrystalline microstructure. Low ductility in ordered alloys can come from a lack of slip system preventing plastic deformation, easy crack propagation along grain boundaries or intrinsic brittle intergranular fracture, segregation, restricted cross slip, difficulty in twinning or impurity locking of dislocations [22].

In this paper, we compare different microstructure and mechanical properties of Ni-Mn-based Heusler as-spun long ribbons in view to have a better understanding of the limiting factors for the use of these materials in elastocaloric devices.

## 2. Experimental details

In this work, the martensitic transformation, microstructure, and mechanical property of $Ni_{(50-x)}Fe_xMn_{32}Al_{18}$ (x = 4, 5, 6) (at. %) and NiX(X = Fe, Co)MnIn (nominal composition $Ni_{50}Fe_2Mn_{34}In_{14}$ and $Ni_{45}Co_5Mn_{36.9}In_{13.1}$ (at. %)) as-spun ribbons were studied. To prepare the as-spun ribbon, an ingot with a weight of 20 g was first prepared by induction heating high-purity constituent elements. The ingot was remelted 4 times to ensure compositional homogeneity. Then the ingot was melt spun around 1523 K into polycrystalline ribbons with a dimension of 20 – 50 μm thickness, 2.5 – 3.5 mm width and 20 – 300 mm length. The tangential speed of the rotating wheel for melt spinning is 18 m·s$^{-1}$. Both of the induction heating and the melt spinning were carried out under high purity argon atmosphere. **Fig. 1** is a schematic diagram of melt spinning and the geometry of the as-spun ribbon. For the convenience of description, the surface of the as-spun ribbon, which solidified against the rotating copper wheel, is defined as the wheel side surface; the opposite one, is defined as the air side surface; the direction perpendicular to the air side surface is defined as the thickness direction (TD); the direction along the longitude direction of the as-spun ribbon is defined as the spinning direction (SD); and the direction perpendicular to TD and SD is defined as the width direction (WD).

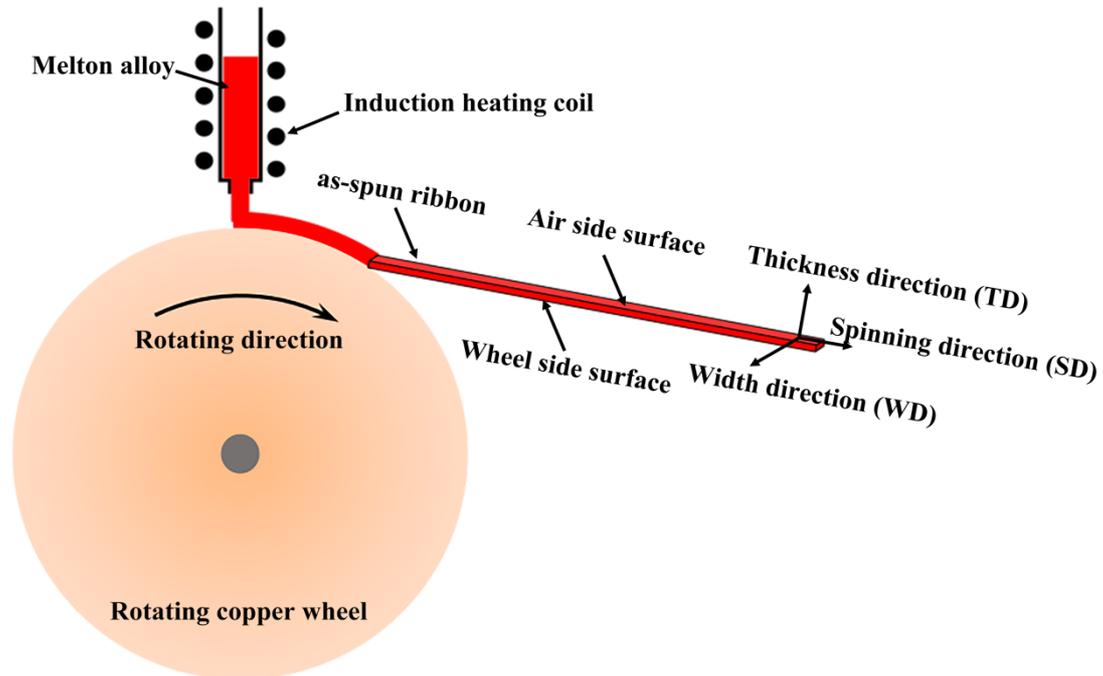

**Fig. 1**. Schematic diagram of melt spinning and the geometry of the as-spun ribbon.

The thermal and magnetic signals during the forward and reverse martensitic transformation were recorded by differential scanning calorimetry (DSC, TA-Q200) with a heating/cooling rate of 10 K·min$^{-1}$ under nitrogen atmosphere and a homemade magnetometer, respectively. Crystal structure was determined by X-ray diffraction (XRD) (PANalytical X'Pert PRO MPD) using Cu Kα (λ = 1.540598Å) radiation at room temperature.

Microstructural observations, chemical composition analysis and crystallographic characterizations were performed by scanning electron microscopes (SEM, Zeiss Ultra55 and Zeiss Gemini SEM 500) equipped with a silicon drift detector (QUANTAX EDS, Bruker) and a Hikari Pro EBSD camera on the longitude section of the as-spun ribbons. To prepare this sample, the ribbons were sandwiched along the SD between two pieces of Si substrate using Epotek G1 glue, and polished down with a final silica colloid step to minimize surface roughness.

Nanoindentation experiments (MTS-XP) were used to extract the mechanical properties (indentation elastic modulus and true hardness) on longitude section of ribbons (using the same samples for SEM-EBSD studies) using a diamond Berkovich tip geometry, following the standard Oliver and Pharr procedure [26]. Continuous Stiffness method was used with an imposed strain rate of 0.05 s$^{-1}$ and a superimposed 2 nm oscillation at 40 Hz to a maximum penetration depth of 300 nm. Matrix of indents (8 × 10 positions with a 10 μm × 5 μm pitch) were carried out to map modulus and hardness throughout the whole thickness of the ribbons (between 30 – 50 μm thickness). Furthermore, the mechanical properties of the as-spun ribbons were studied by bending test on a 3-points bending device (Gatan Microtest$^{TM}$ 300B). A digital camera was connected to a binocular microscope to record the ribbon profile with a frequency of one image per second during the bending test until fracture. The imposed displacement rate of the bending test was set around 8 μm·s$^{-1}$.

## 3. Results and discussion

### 3.1 Transition temperature characterization

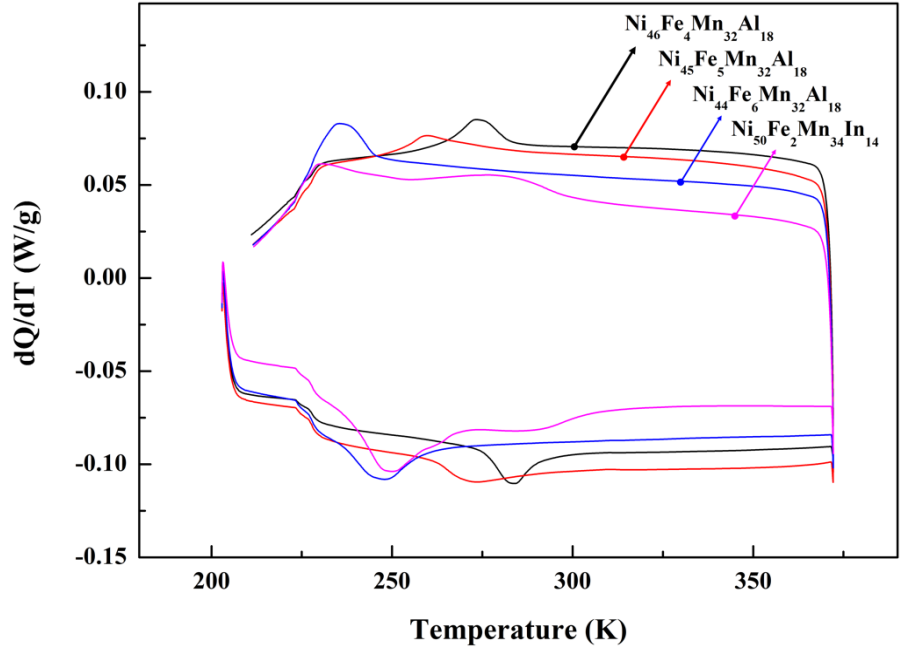

**Fig. 2**. Cooling and heating DSC curves of $Ni_{(50-x)}Fe_xMn_{32}Al_{18}$ (x = 4, 5, 6) and $Ni_{50}Fe_2Mn_{34}In_{14}$ as-spun ribbon.

**Fig. 2** displays the heating and cooling DSC curves of $Ni_{(50-x)}Fe_xMn_{32}Al_{18}$ (x = 4, 5, 6) (at. %) and $Ni_{50}Fe_2Mn_{34}In_{14}$ (at. %) as-spun ribbons. The exothermic and endothermic peaks on cooling and heating indicate the first order phase transformations (martensitic and its reverse transformation). The martensitic transformation start temperature $M_s$, martensitic transformation finish temperature $M_f$, austenitic transformation start temperature $A_s$ and austenitic transformation finish temperature $A_f$ determined using the tangent method. The forward and inverse martensitic transformation entropy change, $\Delta S_{A \to M}$ and $\Delta S_{M \to A}$, are also determined based on the latent heat and the DSC curves. All these results are summarized in **Table 1**. Clearly, for $Ni_{(50-x)}Fe_xMn_{32}Al_{18}$ (x = 4, 5, 6) as-spun ribbons, the martensitic transformation temperatures decrease with the increase of Fe content. The $M_s$ and $A_f$ temperatures of $Ni_{46}Fe_4Mn_{32}Al_{18}$ ribbon is around RT, which means the martensite phase and austenite phase probably co-exist at RT. By doping more Fe, the phase transformation temperatures decrease. It indicates that the other two Ni-Fe-Mn-Al as-spun ribbons are in single austenite phase at RT. To ensure comparison and reproducibility of results, certain heat treatment were conducted on $Ni_{46}Fe_4Mn_{32}Al_{18}$ to acquire a single austenite state (for microstructure and mechanical properties studies). The phase transition entropy change for both Ni-Fe-Mn-Al and Ni-Fe-Mn-In compositions are in the same range. So the refrigeration compatibility, that resulting from the elastocaloric effect, is roughly the same. To be noticed, due to instrumental base line uncertainty and instrumental measurement range limit, the $M_f$ temperature and forward martensitic transformation entropy change ($\Delta S_{M \to A}$) of $Ni_{44}Fe_6Mn_{32}Al_{18}$ and $Ni_{50}Fe_2Mn_{34}In_{14}$ as-spun ribbons cannot be determined accurately from the DSC curves.

**Table 1**. Martensitic transformation start and finish temperatures, and austenitic transformation start, finish temperatures and forward ($\Delta S_{M \to A}$) and inverse ($\Delta S_{A \to M}$) martensitic transformation entropy change of $Ni_{(50-x)}Fe_xMn_{32}Al_{18}$ (x = 4, 5, 6) (at. %) and $Ni_{50}Fe_2Mn_{34}In_{14}$ (at. %) as-spun ribbons determined from DSC data. (*Data with uncertainty due to instrumental base line and measurement range limit.)

| Composition | $M_s$ / K | $M_f$ / K | $A_s$ / K | $A_f$ / K | $\Delta S_{M \to A}$ / J·kg$^{-1}$·K$^{-1}$ | $\Delta S_{A \to M}$ / J·kg$^{-1}$·K$^{-1}$ |
|---|---|---|---|---|---|---|
| $Ni_{46}Fe_4Mn_{32}Al_{18}$ | 282 | 264 | 275 | 290 | 24 | 23 |
| $Ni_{45}Fe_5Mn_{32}Al_{18}$ | 275 | 251 | 261 | 279 | 12 | 10 |
| $Ni_{44}Fe_6Mn_{32}Al_{18}$ | 245 | 224* | 237 | 259 | 19 | 35* |
| $Ni_{50}Fe_2Mn_{34}In_{14}$ | 239 | 222* | 240 | 255 | 22 | 9* |

In order to understand the magnetic properties and the effect of Fe addition, the magnetization was recorded as a function of temperature (M-T) for $Ni_{(50-x)}Fe_xMn_{32}Al_{18}$ (x = 4, 5, 6) and $Ni_{50}Fe_2Mn_{34}In_{14}$ as-spun ribbons under a magnetic field of 7 T, as shown in **Fig. 3(a) – (d)**. From the field cooling (FC) and field cooled heating (FCH) M-T curves of $Ni_{(50-x)}Fe_xMn_{32}Al_{18}$ (x = 4, 5, 6) as-spun ribbons (**Fig. 3(a) – (c)**), it can be seen that, during the cooling-heating cycle, all the ribbons undergo a first order phase transition between a weak ferromagnetic martensite phase and a weak ferromagnetic austenite phase, accompanying a variation of magnetization and a thermal hysteresis. The magnetization difference between martensite phase and austenite phase increase from 1.9 emu·g$^{-1}$ (x = 4) to 6.2 emu·g$^{-1}$ (x = 6) at 7 T external magnetic field and the phase transition temperatures shift to lower temperature with the increase of Fe content. Different from the $Ni_{(50-x)}Fe_xMn_{32}Al_{18}$ (x = 4, 5, 6) as-spun ribbons (shown in **Fig. 3(d)**), with the decreasing of temperature, the $Ni_{50}Fe_2Mn_{34}In_{14}$ one first undergoes a second order magnetic transition from paramagnetic austenite to ferromagnetic austenite and then a first order phase transition from ferromagnetic austenite to ferromagnetic martensite. It can be seen that, at low temperature the magnetization of $Ni_{50}Fe_2Mn_{34}In_{14}$ as-spun ribbon after 7 T magnetic field cooled is twice higher than zero field cooled.

This originates from the kinetic arrest effect in metamagnetic Ni-Mn-based SMAs [27, 28]. During the field cooling process, the high external magnetic field can stabilize the high magnetization phase. Hence, after high external magnetic field cooling, austenite and martensite phases coexist in the ribbon. This effect was not observed in $Ni_{(50-x)}Fe_xMn_{32}Al_{18}$ (x = 4, 5, 6) as-spun ribbons, since it possesses a tiny magnetization difference between the austenite and martensite phase (less than 10% of $Ni_{50}Fe_2Mn_{34}In_{14}$ as-spun ribbon). As is known, the magnetic entropy change plays a negative role on the transformation entropy change in metamagnetic Ni-Mn-based SMAs, a larger magnetic entropy change is adverse for elastocaloric effect. In other words, materials like Ni-Mn-Al-based SMAs, which possesses a negligible magnetic entropy change, could be potential candidates for elastocaloric application.

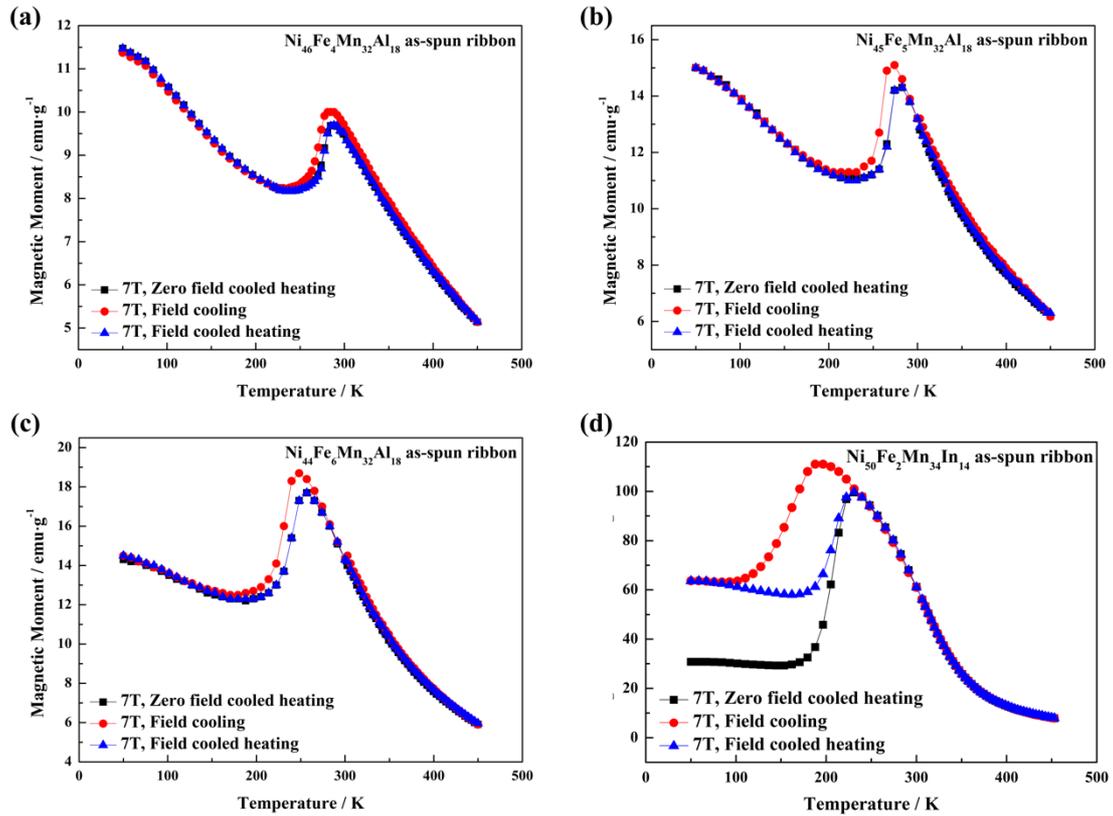

**Fig. 3** Zero field cooled heating (ZFCH), field cooling (FC) and field cooled heating (FCH) thermo-magnetization (M-T) curves obtained for (a) $Ni_{46}Fe_4Mn_{32}Al_{18}$, (b) $Ni_{45}Fe_5Mn_{32}Al_{18}$, (c) $Ni_{44}Fe_6Mn_{32}Al_{18}$ and (d) $Ni_{50}Fe_2Mn_{34}In_{14}$ as-spun ribbons.

### 3.2 Microstructure and crystallographic characterization

**Fig. 4(a) – (d)** shows the typical RT longitude section SEM Backscattered Electron (BSE) micrographs of $Ni_{(50-x)}Fe_xMn_{32}Al_{18}$ (x = 4, 5, 6) and $Ni_{50}Fe_2Mn_{34}In_{14}$ as-spun ribbons. From the micrographs, it can be seen that the wheel side is flat, while the air side has some roughness. Global SEM-BSE observation shows the thickness of all the ribbon varies from 20 μm to 60 μm. As shown in **Fig. 4(a) – (c)**, the $Ni_{(50-x)}Fe_xMn_{32}Al_{18}$ (x = 4, 5, 6) as-spun ribbons is composed of austenite phase (matrix phase, indicated with yellow arrow in the micrograph) and secondary γ phase (indicated with blue arrows in the micrographs). Chemical composition analysis was conducted on Ni-Fe-Mn-Al ribbons. Electron diffraction spectroscopy (EDS) results show that the matrix composition of $Ni_{(50-x)}Fe_xMn_{32}Al_{18}$ (x = 4, 5, 6) ribbons are $Ni_{45.60}Fe_{4.56}Mn_{33.59}Al_{16.25}$ (x = 4), $Ni_{44.82}Fe_{5.72}Mn_{33.61}Al_{15.85}$ (x = 5) and $Ni_{42.71}Fe_{6.66}Mn_{31.88}Al_{18.75}$ (x = 6), respectively. However, due to the dimension of the secondary phase (less than 1 μm in width) and the resolution of the EDS detector (volume around $1 \times 1 \times 1$ μm$^3$), the composition of the secondary phase can not be determined by EDS detector. During the solidification process, due to the high undercooling, the liquid metal first solidified into a thin layer of fine equiaxed austenite grains along the wheel side surface. The grain size of these equiaxed austenite grains is around 1 μm. Then, elongated grains start to generate and grow along the temperature gradient direction, *i.e.* perpendicular to the surface of the ribbon. Previous study on polycrystalline Ni-Fe-Mn-Al alloys with similar compositions reveals that, the secondary phase first appears in $Ni_{47}Fe_3Mn_{32}Al_{18}$ alloy and its amount increases by doping more Fe element [9]. Hence, in our case of higher Fe content, during the solidification the solute atoms redistribute and generate the secondary phase after the Fe element in liquid phase reaches its saturation. It is seen that, the secondary phase generates from the elongated grains layer and ends until the air side surface. Some of the secondary phase distributes continually along the grain boundary of the austenite phase and the other distributes dispersedly in the austenite grains. During the melt spinning process, to reach the liquid metal viscosity for injection, the Ni-Fe-Mn-Al compositions need to be heated to a much higher temperature than the Ni-Mn-In-based compositions, which is closed to the melting temperature of the quartz crucible under the same injection pressure. This indicates that, the molten Ni-Fe-Mn-Al compositions have a higher viscosity. This special intrinsic property and the existence of the secondary phase might lead to a constitutional supercooling during the solidification. It directly results in the formation of large equiaxed austenite grain layer (grain size between 10 μm – 30 μm) along the air side surface of the ribbon. Different from the $Ni_{(50-x)}Fe_xMn_{32}Al_{18}$ (x = 4, 5, 6) as spun ribbons, the $Ni_{50}Fe_2Mn_{34}In_{14}$ as-spun ribbon is composed of pure austenite phase, as shown in **Fig. 4(d)**. The austenite phase has only two morphologies, a layer of fine equiaxed austenite grains along the wheel side surface and the elongated austenite grains that generate from the interface between the equiaxed and elongated grains until the air side surface. The formation mechanisms of the fine equiaxed grains and the elongated grains in $Ni_{50}Fe_2Mn_{34}In_{14}$ as-spun ribbon is the same as the $Ni_{(50-x)}Fe_xMn_{32}Al_{18}$ (x = 4, 5, 6) ones. Clearly, the mobility of the liquid $Ni_{50}Fe_2Mn_{34}In_{14}$ metal is much better and there is no secondary phase generates during the solidification. Hence, the large equiaxed austenite grains did not form during solidification. As is known, the anisotropy of equiaxed grains performs better mechanical properties than the isotropic elongated grains, as well as the dispersed secondary phase. Hence, by optimizing the composition of the ribbon and the resulted microstructures may enhance the mechanical properties of these Ni-Mn-based as-spun ribbons, especially for the application of the elastocaloric effect.

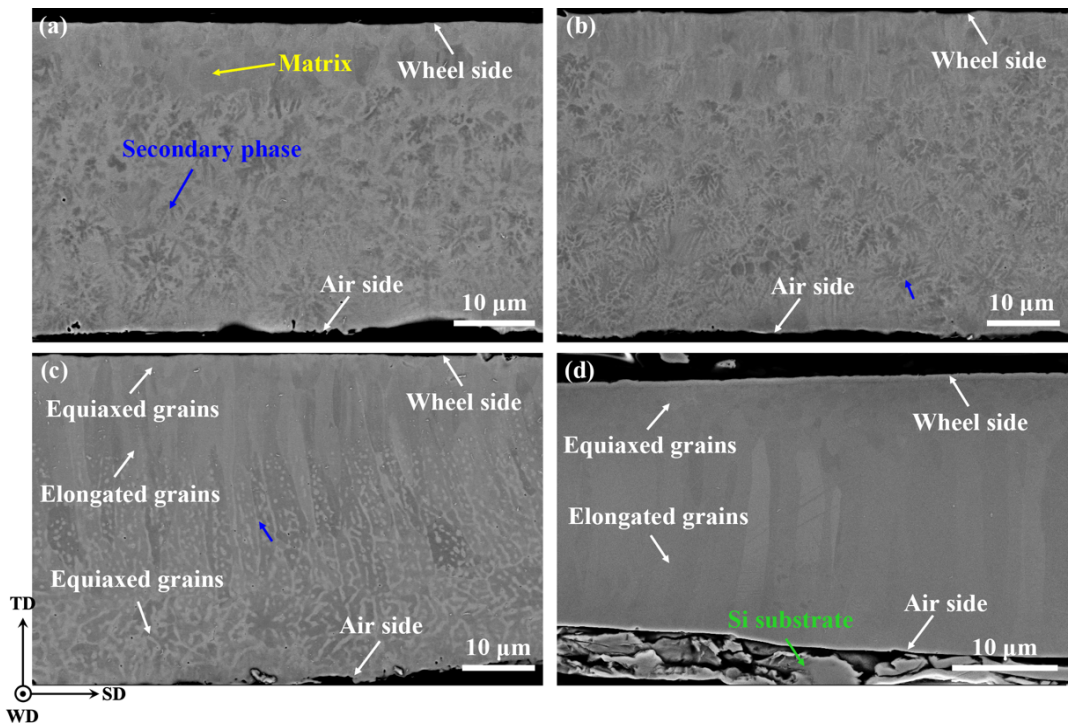

**Fig. 4**. Longitude section SEM Backscattered Electron (BSE) micrographs of (a) – (c) $Ni_{(50-x)}Fe_xMn_{32}Al_{18}$ (x = 4, 5, 6) and (d) $Ni_{50}Fe_2Mn_{34}In_{14}$ as-spun ribbons.

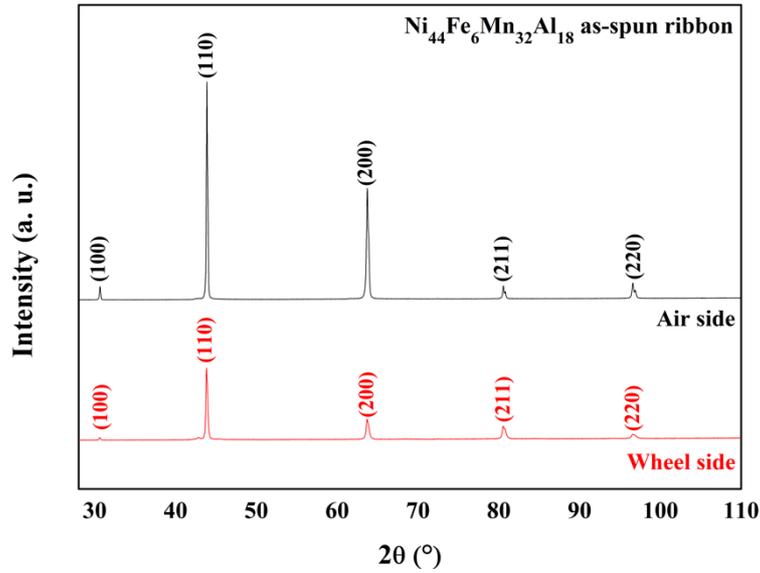

**Fig. 5** XRD patterns of Ni$_{44}$Fe$_6$Mn$_{32}$Al$_{18}$ as-spun ribbon measured at room temperature.

RT XRD pattern of Ni$_{44}$Fe$_6$Mn$_{32}$Al$_{18}$ as-spun ribbon is shown in **Fig. 5**. Clearly, this ribbon is composed of a B2 austenite phase and a cubic secondary phase, which has been demonstrated by previous SEM-BSE observation (**Fig. 4**). The lattice constant of the austenite phase is $a = b = c = 2.9186$ Å, $\alpha = \beta = \gamma = 90°$ (computed with the wheel side surface X-ray diffraction pattern). For Ni$_{50}$Fe$_2$Mn$_{34}$In$_{14}$ as-spun ribbon, the austenite phase also possesses a B2 structure, which has been demonstrated by previous study of Wu *et al.* [29] on a Ni-Fe-Mn-In alloy with similar composition. The lattice parameter of this B2 austenite is $a = b = c = 2.995$ Å, $\alpha = \beta = \gamma = 90°$ [29]. With this crystal structure information, further crystallographic characterization was conducted by SEM-EBSD technique.

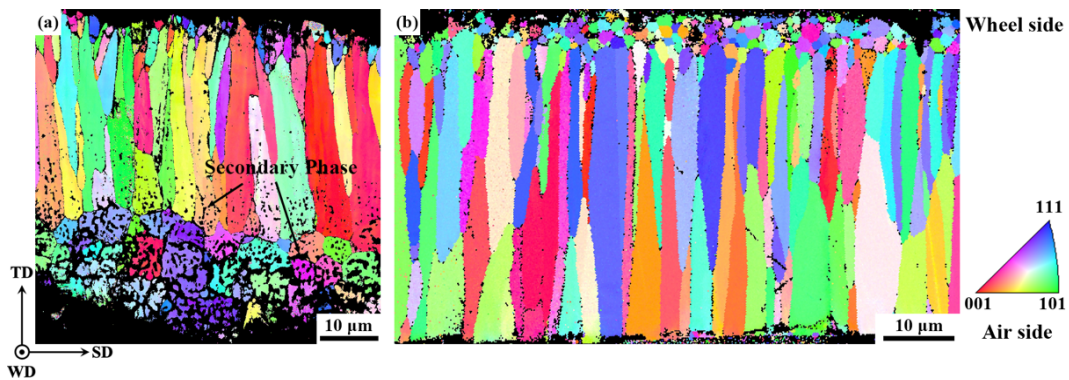

**Fig. 6** Austenite phase EBSD Inverse Pole Figures (IPFs) of (a) Ni$_{44}$Fe$_6$Mn$_{32}$Al$_{18}$ and (b) Ni$_{50}$Fe$_2$Mn$_{34}$In$_{14}$ as-spun ribbon on longitude section. The IPFs are indexed with single austenite crystal structure. The black un-indexed phase in (a) is the secondary γ phase.

The EBSD Inverse Pole Figure (IPF) of $Ni_{44}Fe_6Mn_{32}Al_{18}$ and $Ni_{50}Fe_2Mn_{34}In_{14}$ austenite phase (longitude section) are displayed in **Fig. 6(a)** and **6(b)**, respectively. The longitude section IPF image (**Fig. 6(a)**) of $Ni_{44}Fe_6Mn_{32}Al_{18}$ as-spun ribbon clearly shows that, fine equiaxed grains, elongated grains and large equiaxed grains distribute from wheel side to air side of the ribbon successively. The area in black is the secondary γ phase, since the IPF is indexed with single austenite crystal structure. It is seen that, a small amount of the secondary phase dispersedly distributes in the elongated grains and the other distributes in the large equiaxed grains. While in $Ni_{50}Fe_2Mn_{34}In_{14}$ ribbon, fine equiaxed grains distribute along the wheel side surface and elongated grains along the air side. All these morphological features are in good accordance with previous SEM-BSE observations.

The texture of the austenite grains is further studied. **Fig. 7(a) – (c)** shows the <001>, <111> and <110> pole figures of large equiaxed, elongated, and fine equiaxed austenite grains in $Ni_{44}Fe_6Mn_{32}Al_{18}$ as-spun ribbon, which are calculated based on a set of EBSD orientation data. Clearly, both of the elongated grains and the fine equiaxed grains possess a strong {001} fiber texture (<001>//TD). For the large equiaxed grains, although the <001> direction is more randomly oriented than the elongated grains and the fine equiaxed grains, but it still possesses an {001} fiber texture. And for $Ni_{50}Fe_2Mn_{34}In_{14}$ as-spun ribbon, the elongated grains and the fine equiaxed grains also possess a strong {001} fiber texture. This texture should originate from the temperature gradient during the solidification process. Similar texture was also detected in directional solidified bulk samples in other Ni-Mn-based alloys [30].

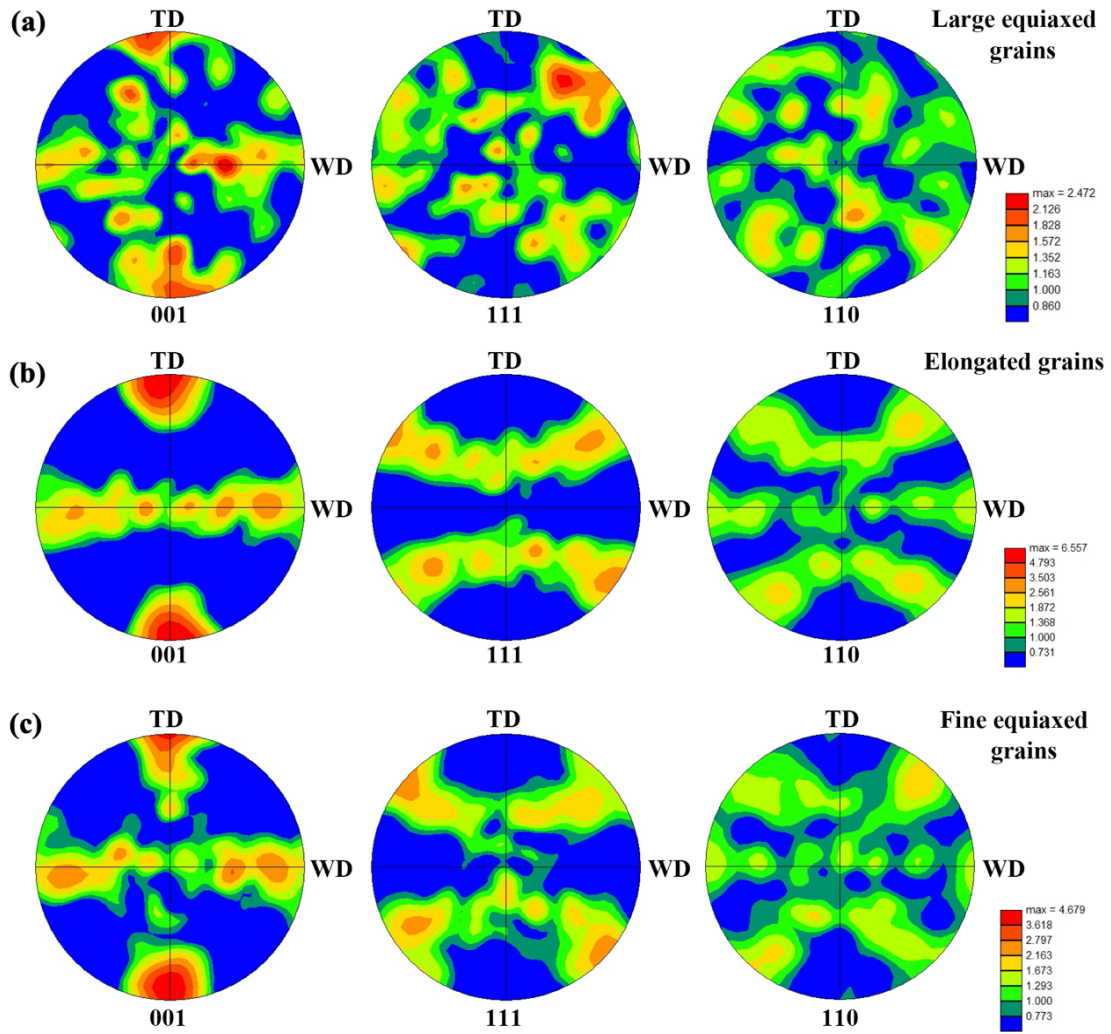

**Fig. 7** <001>, <111> and <110> pole figures of (a) large equiaxed, (b) elongated, and (c) fine equiaxed austenite grains in $Ni_{44}Fe_6Mn_{32}Al_{18}$ as-spun ribbon.

### 3.3 Mechanical property

Mechanical properties are characterized by nanoindentation and 3-points bending test on the various compositions.

### 3.3.1 Nanoindentation

**Fig. 8** shows the cartography of indents along with the hardness profile averaged through the thickness. For all studied compositions, hardness across the thickness varies within 10 %, a variation somehow not correlated with the microstructure change between the wheel / air side depicted above. Since the size of the nanoindents (100 nm depth corresponding to a ~ 1 μm³ plastic volume) are within the scale of the smallest grain diameter, no Hall-Petch effect is perceived. The indentation elastic moduli and hardness of all compositions are in the same range as reported in the literature ([31, 32] for NiMnIn type alloys and [33, 34] for NiMnGa alloys), namely in the 120 GPa and between 4.5 – 6 GPa respectively for indentation modulus and hardness, as shown in **Fig. 9**.

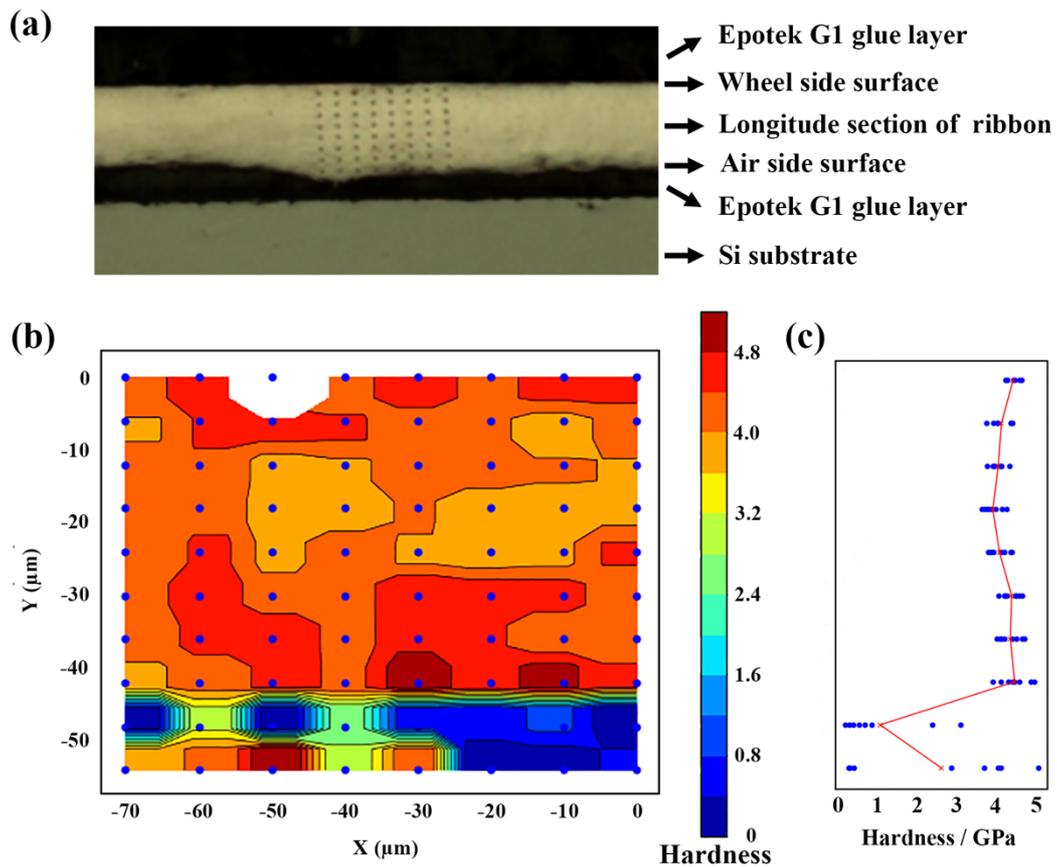

**Fig. 8** (a) Optical micrograph of the nanoindents array (10 × 5 μm pitch), (b) corresponding hardness cartography (100 nm depth), (c) hardness profile across TD.

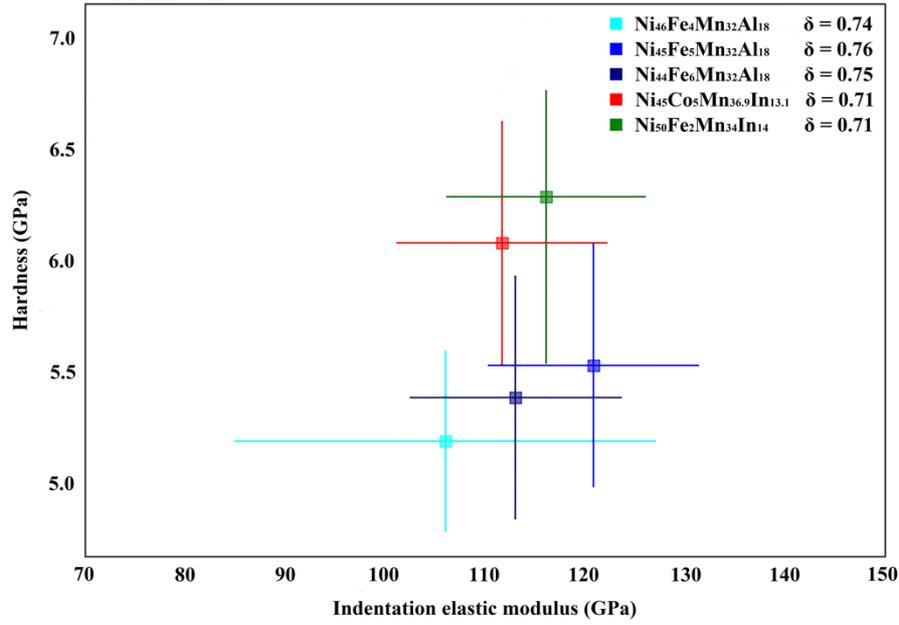

**Fig. 9** Hardness (100 nm depth) versus Indentation elastic modulus for various alloys. Corresponding ductility index ($\delta$) is given in the legend (see text for explanation).

It is possible to extract a ductility index, the dimensionless parameter $\delta$, from indentation measurements satisfying the physical definition of plasticity [35, 36]:

$$\delta = \varepsilon_p/\varepsilon_t = 1 - \varepsilon_e/\varepsilon_t$$

where $\varepsilon_p$, $\varepsilon_t$ and $\varepsilon_e$ are the plastic, total and elastic strain respectively. Using an improved Johnson inclusion model of indentation by pyramidal indenter (Berkovich), [35, 36] provides an analytical formulation of $\delta$ based on indentation elastic modulus and hardness. $\delta$ values are reported in the legend of **Fig. 9** for the studied compositions: Ni-Fe-Mn-Al compositions show an improved ductility ($\delta = 0.75$) as compared with Ni-Fe-Mn-In ($\delta = 0.71$) compositions and $Ni_{45}Co_5Mn_{50-x}In_x$ ($12.5 \leq x \leq 13.2$) ($\delta = 0.71$) [32]. Despite this slight improvement, the ductility index is equivalent to low-ductility intermetallics compounds, well below typical value for steel alloys ($\delta > 0.9$) or even ductile face-centered cubic metals ($\delta > 0.95$) [36].

### 3.3.2 3-points bending test

Results of the 3-points bending test are discussed qualitatively to compare Al versus In substitution with $Ni_{50}Fe_2Mn_{34}In_{14}$ and $Ni_{44}Fe_6Mn_{32}Al_{18}$ as-spun ribbons, corresponding to the two ductility index obtained in the previous section.

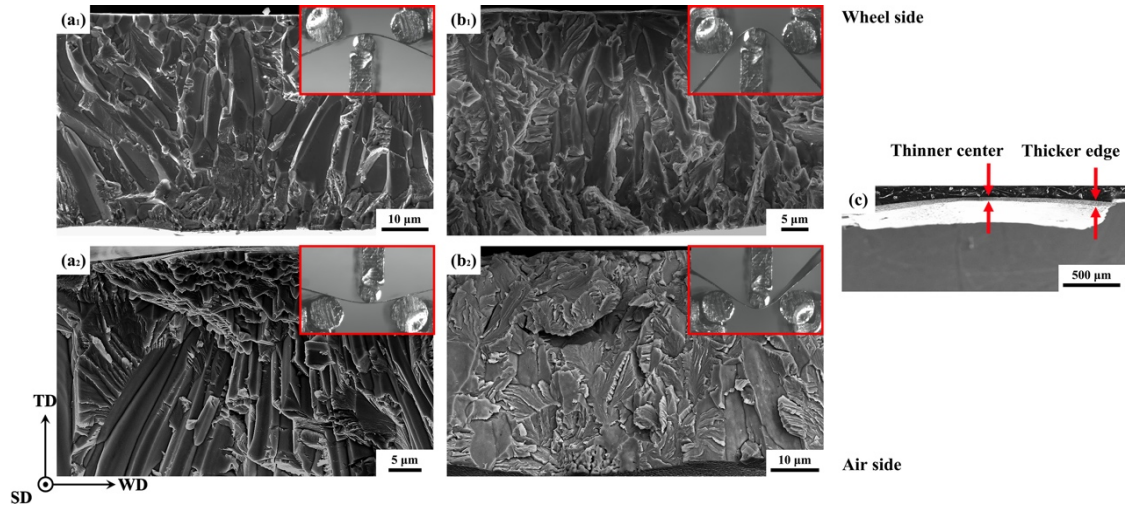

**Fig. 10** Cross section fracture SEM-BSE micrographs (wheel side up, air side down) of (a$_1$) Ni$_{50}$Fe$_2$Mn$_{34}$In$_{14}$ wheel side tensile; (a$_2$) Ni$_{50}$Fe$_2$Mn$_{34}$In$_{14}$ air side tensile; (b$_1$) Ni$_{44}$Fe$_6$Mn$_{32}$Al$_{18}$ wheel side tensile; (b$_2$) Ni$_{44}$Fe$_6$Mn$_{32}$Al$_{18}$ air side tensile bending test, and (c) full scale cross section of Ni$_{44}$Fe$_6$Mn$_{32}$Al$_{18}$ ribbon. The four insets in (a$_1$), (a$_2$), (b$_1$) and (b$_2$) shows the maximum bending displacement before fracture.

**Fig. 10** shows the maximum bending displacement before fracture and the cross section SEM-BSE fractography of Ni$_{44}$Fe$_6$Mn$_{32}$Al$_{18}$ and Ni$_{50}$Fe$_2$Mn$_{34}$In$_{14}$ as-spun ribbons under wheel side tensile and air side tensile state. It can be seen that, all fractographies show mainly an intergranular fracture process both in elongated and equiaxed grains (in **Fig. 10(a$_1$)**, **(a$_2$)**, **(b$_1$)** and **(b$_2$)**). The insets in the fractographies show that, for both compositions, the wheel side in tension displays a larger bending displacement. This is noticeable for the Ni-Mn-In-based ribbons, where the wheel side can be bent up to 2 times strain of the air side. This difference in behavior is less marked for the Ni-Fe-Mn-Al ribbons, with a difference around 20 %. We interpret this difference being due to the microstructure (see **Fig. 6**): indeed equiaxed grains on both air and wheel side could resist better to tension than long elongated grains on the air side.

Using the ribbon profile and the measured thickness on fracture, the maximum strain before fracture can be determined. **Table 2** shows the determined upper bound estimates of strain for both compositions under wheel and air side tensile states. Comparing the two compositions, the bending displacement of Ni-Fe-Mn-Al ribbons is at least 3 times larger than Ni-Mn-In-based ribbons. Due to the processing parameters, the ribbon always possesses a thicker edge, comparing to the center on cross section, as shown in **Fig. 10(c)**, *i.e.* the thickness is not homogeneous along WD. Hence, precise analysis of the strain is difficult. This cross section shape among different compositions, *i.e.* the width and the ratio of the maximum and minimum thickness of the ribbons, are the same, which allows the comparison.

**Table 2**. Upper bound strain of $Ni_{50}Fe_2Mn_{34}In_{14}$ and $Ni_{44}Fe_6Mn_{32}Al_{18}$ as-spun ribbons under wheel and air side tensile states.

| Composition | Stress state | Upper bound strain |
|---|---|---|
| $Ni_{50}Fe_2Mn_{34}In_{14}$ | Wheel side in tension | 0.9 % |
| | Air side in tension | 0.5 % |
| $Ni_{44}Fe_6Mn_{32}Al_{18}$ | Wheel side in tension | 3.2 % |
| | Air side in tension | 2.1 % |

## 4. Summary

In this work, the microstructure and mechanical properties of $Ni_{(50-x)}Fe_xMn_{32}Al_{18}$ (x = 4, 5, 6) (at. %) and NiX(X = Fe, Co)MnIn (nominal composition $Ni_{50}Fe_2Mn_{34}In_{14}$ and $Ni_{45}Co_5Mn_{36.9}In_{13.1}$ (at. %)) have been investigated in detail. The melt spinning process allows the production of very long ribbons up to 300 mm in length. At RT, the $Ni_{46}Fe_4Mn_{32}Al_{18}$ ribbon is in martensite and austenite co-exist state. By increasing the amount of Fe doping, the phase transformation temperatures decrease. And $Ni_{50}Fe_2Mn_{34}In_{14}$ ribbon is in the austenite state at RT. The magnetism of Ni-Fe-Mn-Al ribbons is much weaker than Ni-Fe-Mn-In ribbons. At low temperature, under a high external magnetic field cooling process, kinetic arrest effect was observed only in Ni-Fe-Mn-In ribbons, but not in Ni-Fe-Mn-Al ribbons. A gradient in microstructure exists between the air side surface and the wheel side surface. This is induced by the temperature gradient during fast rate solidification, and resulting in elongated grains and equiaxed grains, respectively. Both elongated and equiaxed grains possess a strong {001} fiber texture (<001>//TD). Little variation of hardness is measured across the different microstructures. The substitution of In by Al allows to increase very slightly the ductility index which can reach 0.75. However, this increase does not allow to exceed the values associated with intermetallic materials with low ductility. A contrast in bending is observed between the wheel side tensile and the air side tensile. Larger bending displacements are realized in the wheel side in tension. The fine equiaxed grains probably resist the tension on wheel side surface. The substitution of In by Al improves the maximum bending strain by a factor of 3. Overall, the maximum strain is limited to a maximum value of 3 % and intergranular fracture remains the main damage mechanism. All these results provide fundamental information on microstructural features and mechanical properties on flexible Ni-Mn-based as-spun ribbons. Optimized compositions exhibit a maximum bending strain of 3 %: within this range, fatigue properties remain to be investigated for practical application of elastocaloric effect.

**Acknowledgements**

This work was supported by the institute Carnot Energies du Futur under the Project '18T12 – COM' (Grant No. 16 CARN 0010 01) and the authors thank Daniel Bourgault for fruitful discussions.